\documentclass[%
 reprint,
superscriptaddress,
 amsmath,amssymb,
 aps,
]{revtex4-2}

\usepackage{graphicx}
\usepackage{dcolumn}
\usepackage{bm}

\usepackage[export]{adjustbox}
\begin{document}

\preprint{APS/123-QED}

\title{Exceptional point induced quantum phase synchronization and entanglement dynamics in mechanically coupled gain-loss oscillators}

\author{Joy Ghosh}%
\email{joyghos@kgpian.iitkgp.ac.in}
\affiliation{%
School of Nanoscience and Technology, IIT Kharagpur
}%
\author{Souvik Mondal}
\affiliation{%
Electronics and Electrical Communication Engineering Department, IIT Kharagpur, West Bengal, 721302, India
}%
\author{Shailendra K. Varshney}
\affiliation{%
Electronics and Electrical Communication Engineering Department, IIT Kharagpur, West Bengal, 721302, India
}%
\affiliation{%
School of Nanoscience and Technology, IIT Kharagpur
}%
\author{Kapil Debnath}%
\email{ kapil.debnath@abdn.ac.uk}
\affiliation{%
Electronics and Electrical Communication Engineering Department, IIT Kharagpur, West Bengal, 721302, India
}%
\affiliation{%
School of Natural and Computing Sciences, University of Aberdeen, Aberdeen AB24 3UE, UK}%

\begin{abstract}
  The optomechanical cavity (OMC) system has been a paradigm in the manifestation of continuous variable quantum information over the past decade. This paper investigates how quantum phase synchronization relates to bipartite Gaussian entanglement in coupled gain-loss mechanical oscillators, where the gain and loss rates are engineered by driving the cavity with blue and red detuned lasers, respectively. We examine the role of exceptional point in a deterministic way of producing self-sustained oscillations that induce robust quantum correlations among quadrature fluctuations of the oscillators. Particularly, steady phase synchronization dynamics along with the entanglement phenomena are observed in the effective weak coupling regime above a critical driving power. These phenomena are further verified by observing the mechanical squeezing and phase space rotations of the Wigner distributions. Additionally, we discuss how the oscillators' frequency mismatches and decoherence due to thermal phonons impact the system dynamics. These findings hold promise for applications in phonon-based quantum communication and information processing.
\end{abstract}

\maketitle


\section{Introduction}

Synchronization is a natural phenomenon widely observed around us, where two or more systems tend to act similarly at the same time. Huygens initially proposed the notion of a synchronized oscillation in the early 17th century during experiments involving mechanical clocks \cite{bhlitem61160}. Since then, it has been found in various processes such as the flashing of fireflies, chemical reactions, neuron networks, heart cells, etc \cite{Pikovsky2001SynchronizationA}. Synchronization in different classical setups was extensively studied in the past, but in the quantum domain, it gained popularity after Mari et al. proposed a measure to compute complete synchronization and phase synchronization for continuous variable systems \cite{PhysRevLett.111.103605}, which has been applied in different fields like cavity QED \cite{LI2017121}, atomic ensembles \cite{PhysRevLett.113.154101}, VdP oscillators \cite{PhysRevA.91.012301}, spin chains \cite{PhysRevLett.121.063601}, etc. The principle of quantum synchronization differs fundamentally from its classical counterpart due to Heisenberg's uncertainty relation \cite{PhysRevLett.111.103605}. Earlier studies have demonstrated that synchronization is closely linked to other quantum correlations such as entanglement \cite{PhysRevLett.121.063601}, mutual information \cite{PhysRevA.91.012301}, and discord \cite{PhysRevA.85.052101}. The coexistence of quantum synchronization and entanglement is a fascinating phenomenon. Previous research has shown that superconducting qubits emitting entangled photons can synchronize \cite{PhysRevLett.121.063601}. Additionally, clock synchronization has been achieved using entangled photons generated through SPDC \cite{quan2016demonstration}. In quantum many-body systems, it has been observed that entanglement and synchronization are closely linked and can lead to collective cooperative behavior \cite{witthaut2017classical}. Moreover, it has been confirmed that spin-1 systems can also be synchronized through entanglement \cite{PhysRevLett.121.063601}.

In this context, optomechanical architectures \cite{aspelmeyer2014cavity} appeared to be a promising platform to test the spontaneous synchronization among micro or nanomechanical oscillators, where two mechanical or cavity modes can be directly coupled through phonon or photon tunneling. Multiple synchronization schemes have been developed in optomechanics \cite{li2016quantum,kalita2021switching,amitai2017synchronization,ying2014quantum}, among which enhancing the nonlinearity is considered a primary feature. Periodic modulation \cite{du2017synchronization,qiao2018quantum,qiao2020quantum} and quadratic coupling \cite{garg2023quantum} are frequently utilized for this purpose. Recently, a counter-intuitive phenomenon of noise-induced synchronization has been observed \cite{PhysRevLett.129.250601}. Experimental realizations such as self-organized phonon lasers \cite{PhysRevLett.124.053604} and synchronization blockade \cite{PhysRevLett.118.243602} are also reported. 
The curiosity about the interplay between quantum synchronization and entanglement in optomechanical setup is emerging in recent times \cite{wang2014nonlinear,liao2019quantum,garg2023quantum, MANJU2023129039}. Most of the works earlier demonstrated this idea by optical coupling only, mechanical interaction-based design to test quantum synchronization is not well-realized in the current literature. 
The idea of exceptional points (EPs) in mechanically coupled gain-loss structures is a novel tool that evolved rapidly in the past decade \cite{ozdemir2019parity}. EPs refer to fundamental degeneracies in gain-loss cavities or waveguides \cite{makris2008beam,guo2009observation,peng2014parity}, where the system's eigenvalues coalesce and become degenerate. EP-based optomechanical structures have been studied for mass sensing \cite{djorwe2019exceptional}, optomechanically induced transparency \cite{lu2018optomechanically}, sideband generation \cite{PhysRevA.108.023517} offering better controllability and low power threshold requirement. It has also been applied to achieve synchronization and frequency-locking effect in the classical domain \cite{djorwe2018frequency,djorwe2020self}. Operation near exceptional point is utilized to delay sudden death of entanglement \cite{chakraborty2019delayed} and it is also reported that entanglement is greatly enhanced in gain-loss OMC systems \cite{tchodimou2017distant}. Based on the previous literature, it appears that EP has the potential to establish entanglement, so the question we propose is, can it also develop robust quantum synchronization? If non-identical oscillators can be entangled and synchronized simultaneously through EP, this can be applied in the field of quantum communication and information processing.

In this paper, we present a configuration consisting of two mechanically coupled OMCs with symmetrical properties. By employing blue and red detuned lasers to drive the cavities, the gain-loss characteristics of the mechanical oscillators can be manipulated \cite{PhysRevA.108.023517}. This configuration can lead to self-sustained oscillations in both oscillators, which is considered to investigate quantum phase synchronization by employing Mari's criterion \cite{PhysRevLett.111.103605}. The entanglement between the two oscillators is also established simultaneously and estimated by logarithmic negativity \cite{simon2000peres}. Based on the numerical calculations, a rich connection between phase synchronization and entanglement is further clarified. The Wigner distributions are plotted to demonstrate the squeezed and synchronized Gaussian states. Our findings reveal that by taking advantage of the exceptional point, we can switch quickly into limit cycle oscillation that generates the sustainable quantum correlation among quadrature fluctuations of the oscillators. Moreover, the system's parameters such as the mechanical coupling rate or driving field strength can be suitably modified to control quantum phase synchronization and entanglement dynamics in a flexible manner. This approach holds the potential to advance our understanding of the relationship between quantum phase synchronization and entanglement phenomena.
The work is organized as follows. In Sec. II, the quantum Langevin equations are described along with the theoretical model. The numerical simulation of quantum correlation properties is defined in Sec. III by taking the covariance matrix approach. Section IV holds the results and discussion about the possible relationship between these two phenomena, and Sec. V concludes the work.

\section{MODEL AND
CLASSICAL DYNAMICS}
\begin{figure}[htbp]
\includegraphics[width=0.65\linewidth]{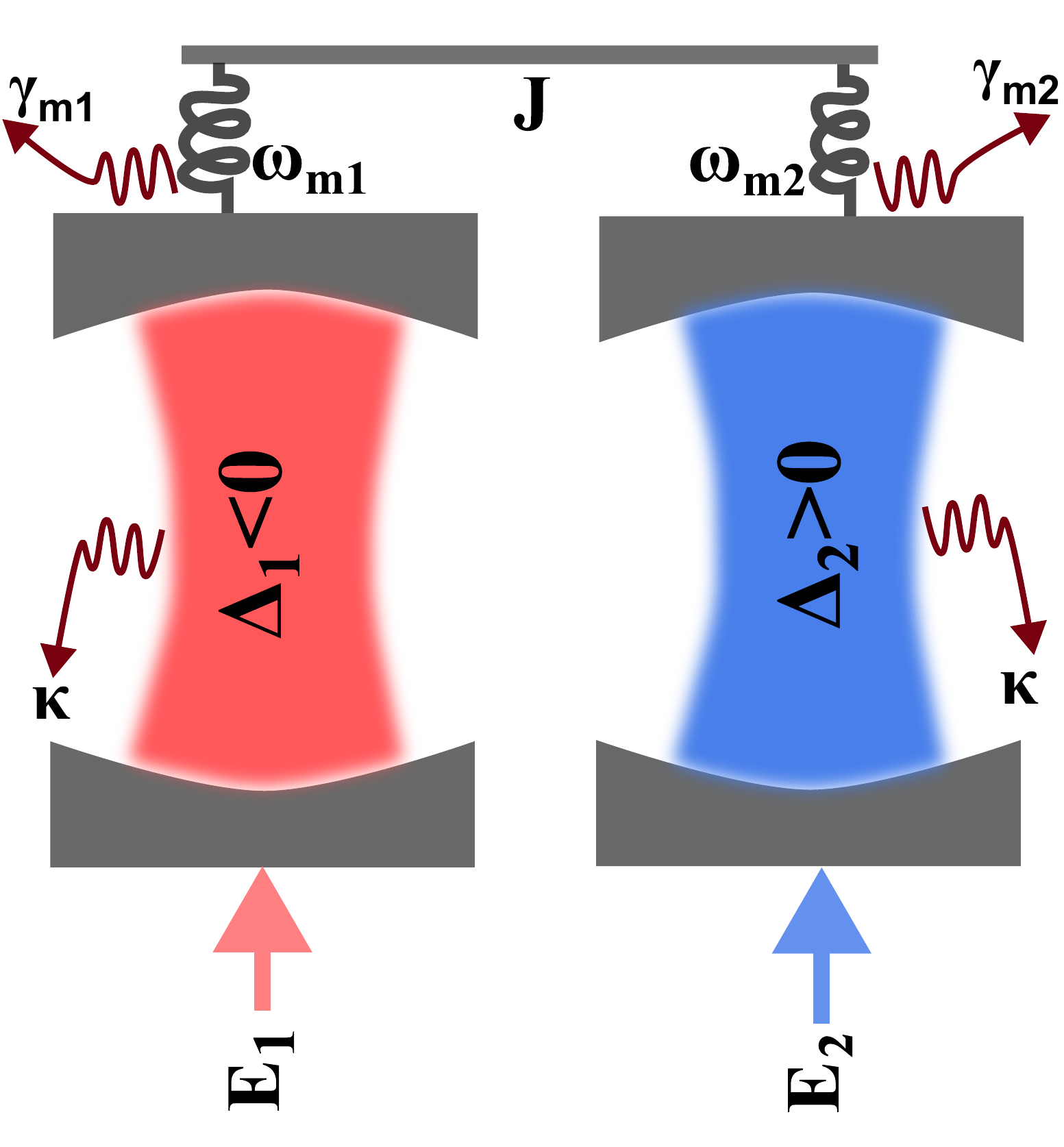}
\caption{\label{fig1} (Color online) Schematic diagram of two Optomechanical cavities coupled mechanically, driven by red ($\Delta_1<0$) and blue ($\Delta_2>0$) detuned laser fields, respectively. The opposite detunings characterize the gain-loss effect, whereas $E_j$ ($j=1,2$) represents the optical driving power incident on one end of the Fabry–Pérot cavity.}
\label{fig:1}
\end{figure}

In this study, we consider a system comprising two optomechanical cavities that are identical but oppositely detuned, both coupled mechanically, as depicted in Fig.\ref{fig:1}. The coupling between the two oscillators is facilitated through phonon tunneling. The Hamiltonian of the complete system can be expressed as follows (taking \(\hbar = 1\))

\begin{eqnarray}
    \mathcal{\hat{H}}&=& \sum_{j=1,2} [-\Delta_{j}\hat{a}^{\dagger}_{j}\hat{a}_{j}+\frac{\omega_{mj}}{2}(\hat{p}_j^2 + \hat{q}_j^2)-g_{0j}\hat{a}^{\dagger}_j\hat{a}_{j}\hat{q}_{j}\nonumber
    \\&&+iE_{j}(\hat{a}^{\dagger}_{j}-\hat{a}_{j})]-J\hat{q}_1\hat{q}_2
    \label{eq:1}
\end{eqnarray}

The Hamiltonian is written here in the rotating frame of the driving frequency ($\omega_L$) with cavity detuning from optical resonance is $\Delta_j=\omega_{oj} - \omega_L$. Here $\hat{a}^{\dagger}_{j}(\hat{a}_{j})$ are the creation(annihilation) operators associated with the optical field with frequency ($\omega_{oj}$) and $\hat{q}_j$ and $\hat{p}_j$ are the dimensionless position and momentum operators of the $j^{th}$ mechanical oscillators with frequencies $\omega_{mj}$. The optomechanical coupling of each cavity is taken as $g_{0j}$ and the laser driving field strength of the two single-mode cavities is $E_j$. The mechanical coupling strength $J$ acts as a bosonic Gaussian channel between the oscillators, assumed to be much smaller than mechanical frequency $(J\ll\omega_j)$. The dissipative dynamics of the system are described by the following set of nonlinear quantum Langevin equations,

\begin{eqnarray}
\partial_t{\hat{a}_j}&=&-(\kappa - i\Delta_j)\hat{a}_j + i g_0\hat{a}_j \hat{q}_j +E_j + \sqrt{2\kappa}\hat{a}_j^{in}
\nonumber\\
\partial_t{\hat{q}_j}&=&\omega_{mj}\hat{p}_j
\nonumber\\
\partial_t{\hat{p}_j}&=&-\omega_{{mj}}\hat{q}_j-\gamma_{mj} \hat{p}_j+J\hat{q}_{3-j}+g_{0}\hat{a}^{\dagger}_{j}\hat{a}_{j} +\hat{\eta}_j
\end{eqnarray}
\label{eq:2}

Here, $\gamma_{mj}$ and $\kappa$ represent the intrinsic dissipation of mechanical oscillators and optical cavities. We have taken the cavity decay rate ($\kappa$) and optomechanical constant ($g_{0j}$) identical for both cavities for simplicity. The laser driving amplitude provided for both cavities is also the same ($E_1=E_2=E$). The stochastic noise operators for optical and mechanical systems are given as $\hat{a}_j^{in}$ and $\hat{\eta}_j$, satisfying the standard correlation 
 $\langle \hat{a}^{in\dagger}_i(t) \hat{a}^{in}_j(t') + \hat{a}_j^{in\dagger}(t')\hat{a}_i^{in}(t)\rangle=\delta_{ij}\delta(t-t')$ and $\frac{1}{2}\langle\hat{\eta}_i(t) \hat{\eta}_j(t') +\hat{\eta}_j(t') \hat{\eta}_i(t)\rangle=\gamma_{mj} (2 \bar{n}_{m} +1)\delta_{ij}\delta(t-t')$ under Markovian approximation \cite{gardiner2004quantum,benguria1981quantum}. The mean thermal phonon occupancy of the mechanical systems at temperature $T$ is taken as same as $\bar{n}_{m}=[\exp(\frac{\hbar\omega_{mj}}{k_B T})-1]^{-1}$ (where $k_B$ is the Boltzmann constant). The quantum Langevin equations are usually solved using the standard linearization technique, where the classical mean dynamics and quadrature fluctuations are separated. The classical dynamical equations, by decomposing cavity and mechanical operators into two parts, $\mathcal{\hat{O}}(t) =\langle \mathcal{\hat{O}}(t)\rangle+\delta\mathcal{O}(t)$, where $\mathcal{O}=a_{j},q_{j},p_{j}$, given as

 \begin{eqnarray}
\partial_t\langle\hat{a}_j\rangle&=&-(\kappa -i\Delta_j)\langle \hat{a}_j\rangle + i g_0 \langle \hat{q}_j\rangle\langle\hat{a}_j\rangle +E 
\nonumber \\ 
\partial_t\langle \hat{q}_j\rangle&=&\omega_{mj} \langle \hat{p}_j\rangle
\nonumber \\ 
\partial_t \langle \hat{p}_j\rangle&=&-\omega_{mj}\langle \hat{q}_j\rangle -\gamma_{mj}\langle \hat{p}_j\rangle +J\langle \hat{q}_{3-j}\rangle
\nonumber\\&&+g_0|\langle\hat{a}_j\rangle|^2
\label{eq:3}
\end{eqnarray}

The linearized equations describing quadrature fluctuations are

\begin{eqnarray}
     \partial_t{\delta a_j}&=&-(\kappa - i\Delta_j)\delta a_j
     + i g_0 (\langle\hat{a}_j\rangle \delta q_j + \langle\hat{a}_j\rangle\delta a_j) \nonumber\\&&+ \sqrt{2\kappa}\hat{a}_j^{in}
    \nonumber\\
    \partial_t{\delta{q_j}}&=&\omega_{mj}\delta{p_j}
    \nonumber\\
    \partial_t{\delta{p_j}}&=&-\omega_{mj}\delta{q_j}-\gamma_{mj}\delta{p_j} +J\delta q_{3-j}
    + g_0(\langle\hat{a}_j\rangle\delta a^{\dagger}_{j}
    \nonumber\\&&
    +\langle\hat{a}^{*}_{j}\rangle\delta a_j)+\hat{\eta}_j
    \label{eq:4}
\end{eqnarray}

Assuming $|\langle a_j\rangle|^2\gg1$, we have ignored all higher-order terms in the above equations. Under this condition, when the cavity decay rate is much larger than the effective optomechanical coupling strength, i.e. $\kappa\gg G_{j}$ (where $G_j=g_0\langle a_j\rangle$), the cavity fields can be safely eliminated from the governing equations \cite{xu2015mechanical, PhysRevA.95.023827}. In that case, the effective Hamiltonian can be written as \cite{PhysRevA.108.023517}
\begin{eqnarray}
    \mathcal{H}_{eff}&=&(\Omega_{m1}-i\Gamma_{m1})b^\dagger_1b_1+(\Omega_{m2}+i\Gamma_{m2})b^\dagger_2b_2\nonumber\\&&
    -J(b_1^\dagger b_2+ H.c)
    \label{eq:5}
\end{eqnarray}
Here the effective mechanical frequency and effective decay (gain) rates are modified, given by $\Omega_{mj}=(\omega_{mj}\pm\Delta \omega_{mj})$ and $\Gamma_{mj}=\gamma_{mj}\mp\gamma_{oj}$ respectively, where $\Delta \omega_{mj}$ and $\gamma_{oj}$ are the optomechanically induced modifications \cite{PhysRevA.108.023517, xu2015mechanical}. In the resolved sideband regime ($\omega_{mj}\gg\kappa$), $\Delta \omega_{mj}$ can be safely neglected and $\gamma_{oj}=4G_j^2/\kappa$. $b_j$ and $b^\dagger_j$ are the creation and annihilation operators related to phonons of the mechanical oscillators, given as $q_j=\frac{(b^\dagger_j+b_j)}{\sqrt{2}}$ and $p_j=\frac{i(b^\dagger_j-b_j)}{\sqrt{2}}$.
The corresponding eigen frequencies of the coupled mechanical modes with $|\Omega_{m1}-\Omega_{m2}|\ll\omega_{mj}$ are given as
\begin{equation}
\begin{split}
\omega_\pm\approx\frac{\Omega_{m1}+\Omega_{m2}}{2}-i\frac{\Gamma_{m1}-\Gamma_{m2}}{4}
\pm \sqrt{J^2 - (\frac{\Gamma_{m1}+\Gamma_{m2}}{4})^2 }
\label{eq:6}
\end{split}
\end{equation}
From Eq.(\ref{eq:6}) it can be seen that the phase transition between strongly coupled and weakly coupled regions occurs at $J=(\Gamma_{m1} + \Gamma_{m2})/2$, which is also known as the exceptional point (EP).
 \begin{figure}[htbp]
\includegraphics[width=\linewidth]{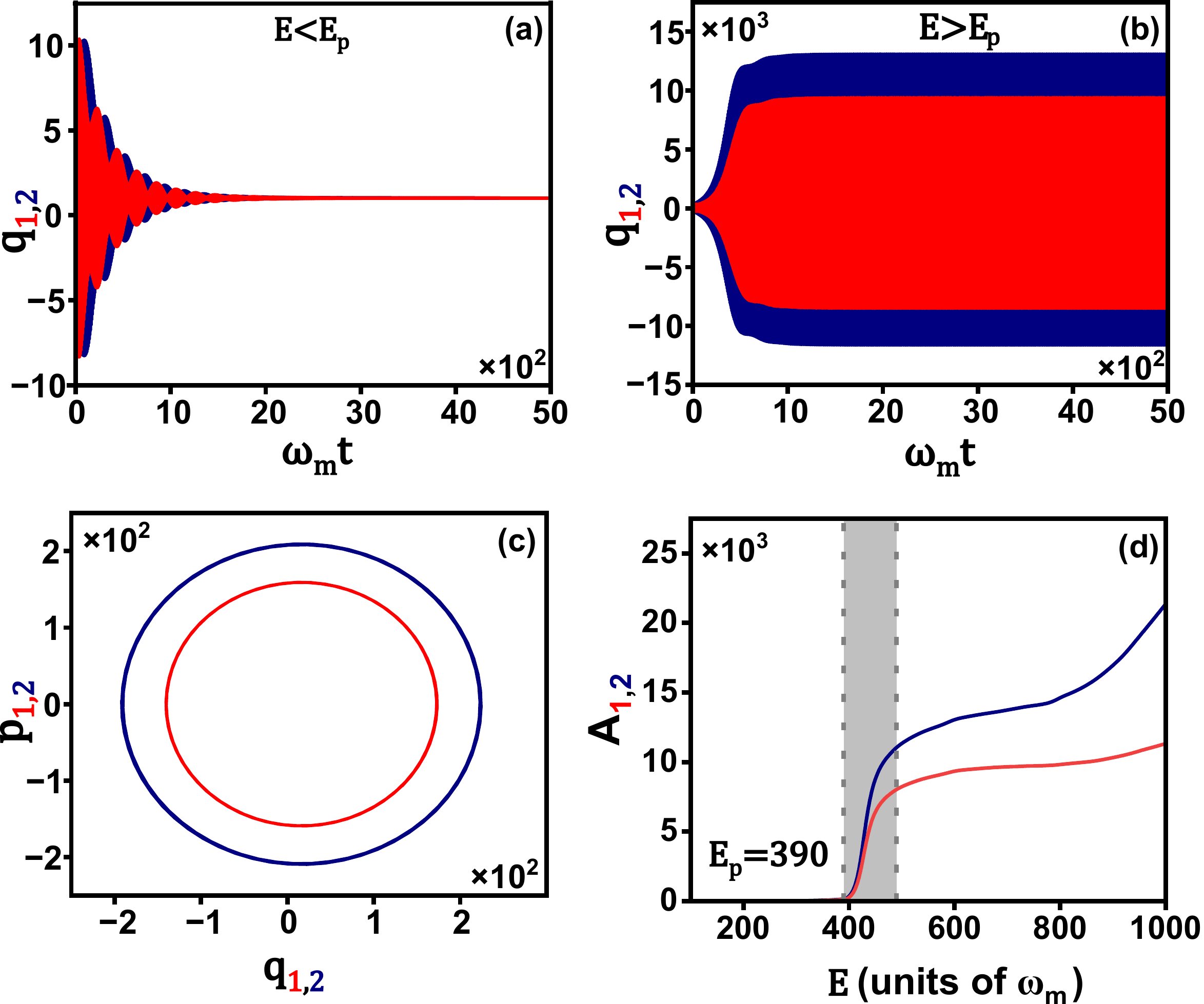}
\caption{\label{fig2}(Color online)
Classical dynamics of (a) decaying oscillation of the mechanical gain (blue) and loss (red) oscillators before reaching EP at driving power level $E=100\omega_m$. (b) Self-sustained oscillation after crossing EP at power level $E=500\omega_m$. (c) Phase-portrait at the same driving strength exhibiting limit cycle. (d) Amplitudes of position variables ($q_{1,2}$) of the mechanical oscillators against driving power $E$, started developing after EP. The shaded region specifies sudden amplification of mechanical amplitudes after the EP which stabilized at $E=490\omega_m$.}
\label{fig:2}
\end{figure}
 We numerically investigate the mechanical dynamics by solving Eq.(\ref{eq:3}) with the following set of parameters, $\omega_{m1}=\omega_m$,  $\omega_{m2}=1.008\omega_m$, $-\Delta_1=\Delta_2=\omega_{m}$, $\kappa= 0.1\omega_{m}$, $\gamma_{m1}=10^{-2}\omega_{m}, \gamma_{m2}= 10^{-4}\omega_{m}$, $g_{0j}= 10^{-4}\omega_{m}$ that can be experimentally achievable in the resolved sideband regime setup. The reason for choosing one mechanical oscillator more dissipative than the other (i.e. $\gamma_{m1}>\gamma_{m2}$) is given in the next section. The ode45 method in Matlab is used to solve the dynamical equations with the initial conditions assumed to be zero for all the variables.
By fixing the mechanical coupling rate at $J=0.03\omega_m (\ll\omega_m)$ and varying the driving amplitude $E$, the classical mechanical dynamics in various regimes can be observed. Fig.\ref{fig:2}(a) depicts the average dynamics of the mechanical positions at driving amplitude, $E=100\omega_m$ which corresponds to the effective coupling condition $J>\frac{\Gamma_{m1}+\Gamma_{m2}}{4}$. The mechanical oscillations show amplitude-modulated dynamics conveying strong coupling between the resonators, but eventually, the dynamics decay with the effective rate $(\Gamma_{m1}-\Gamma_{m2})/4$. Increasing the driving power at $E=500\omega_m$ brings the system in the weakly coupled zone that corresponds to the effective condition $J<\frac{\Gamma_{m1}+\Gamma_{m2}}{4}$. In this regime, the mechanical energies are localized, and the oscillations in both resonators amplify. The optomechanical nonlinearity has saturated the growth as shown in Fig \ref{fig2}(b) with the corresponding phase portrait shown in Fig \ref{fig2}(c). Though we have assumed a small resonance frequency difference, the dynamics in both the mechanical oscillators evolve with a locked phase, consistent with the results shown in \cite{djorwe2019exceptional, PhysRevA.108.023517}. The variation of the mechanical oscillation amplitude $A_{1,2}$ with the driving strength is shown in Fig. \ref{fig:2}(d). From this, it can be found that the transition point occurs at $E=E_p=390\omega_m$. After the EP, a sudden amplification of oscillation (denoted by the shaded region) occurred, and the limit cycles were reached when $E=490\omega_m$. The difference of amplitudes increases with increasing $E$ as the influence of mechanical coupling $J$ becomes weaker, i.e. $J\ll\ (\Gamma_{m1}+\Gamma_{m2})/4$, which results in the effectively lesser flow of mechanical energies from cavity 2 to cavity 1. The transition point is consistent with the analytical solution of the effective mechanical mode picture given by Eq.(\ref{eq:6}).

\section{Quadrature fluctuations and correlations}

In this section, we discuss the quantum correlation properties of the quadrature fluctuations and further define the phase synchronization along with the entanglement generation schemes of the coupled mechanical oscillators. Introducing the quadrature operators for optical fields $\hat{x}_j = \frac{1}{\sqrt{2}}(\delta a^\dagger_j + \delta a_j)$ and $\hat{y}_j = \frac{i}{\sqrt{2}}(\delta a^\dagger_j - \delta a_j)$ and for the noise operators $\hat{x}^{in}_j = \frac{1}{\sqrt{2}}(\delta a^{in\dagger}_j + \delta a^{in}_j)$, $\hat{y}^{in}_j = \frac{i}{\sqrt{2}}(\delta a^{in\dagger}_j - \delta a^{in}_j)$, the set of
Eq.(\ref{eq:4}) describing fluctuations can be rewritten in compact matrix form as
\begin{equation}
\partial_t u = \mathcal{A}(t)u(t) + n(t)
\label{Eq:7}
\end{equation}
Here $u^{T}=(\delta q_1, \delta p_1, \delta x_1, \delta y_1, \delta q_2, \delta p_2, \delta x_2, \delta y_2)$ is the quadrature fluctuation vector and $n^{T} = (0, \eta_1, \sqrt{2\kappa}\delta x^{in}_1, \sqrt{2\kappa}\delta y^{in}_1, 0, \eta_2 ,\sqrt{2\kappa} \delta x^{in}_2,\sqrt{2\kappa} \delta y^{in}_2)$ 
is the input noise vector with the drift matrix $\mathcal{A}$ is given in Appendix A. The statistical correlations of quadrature fluctuations can be found by studying the evolution of the drift matrix $\mathcal{A}$. The formal solution of Eq.(\ref{Eq:7}) can be written as $u(t)=M(t)u(0)+\int_{0}^{t}M(\tau)\mathcal{N}(t-\tau)d\tau$, where $M(t)=e^{\mathcal{A}t}$. 

\begin{figure}[htbp]
\includegraphics[width=0.8\linewidth]{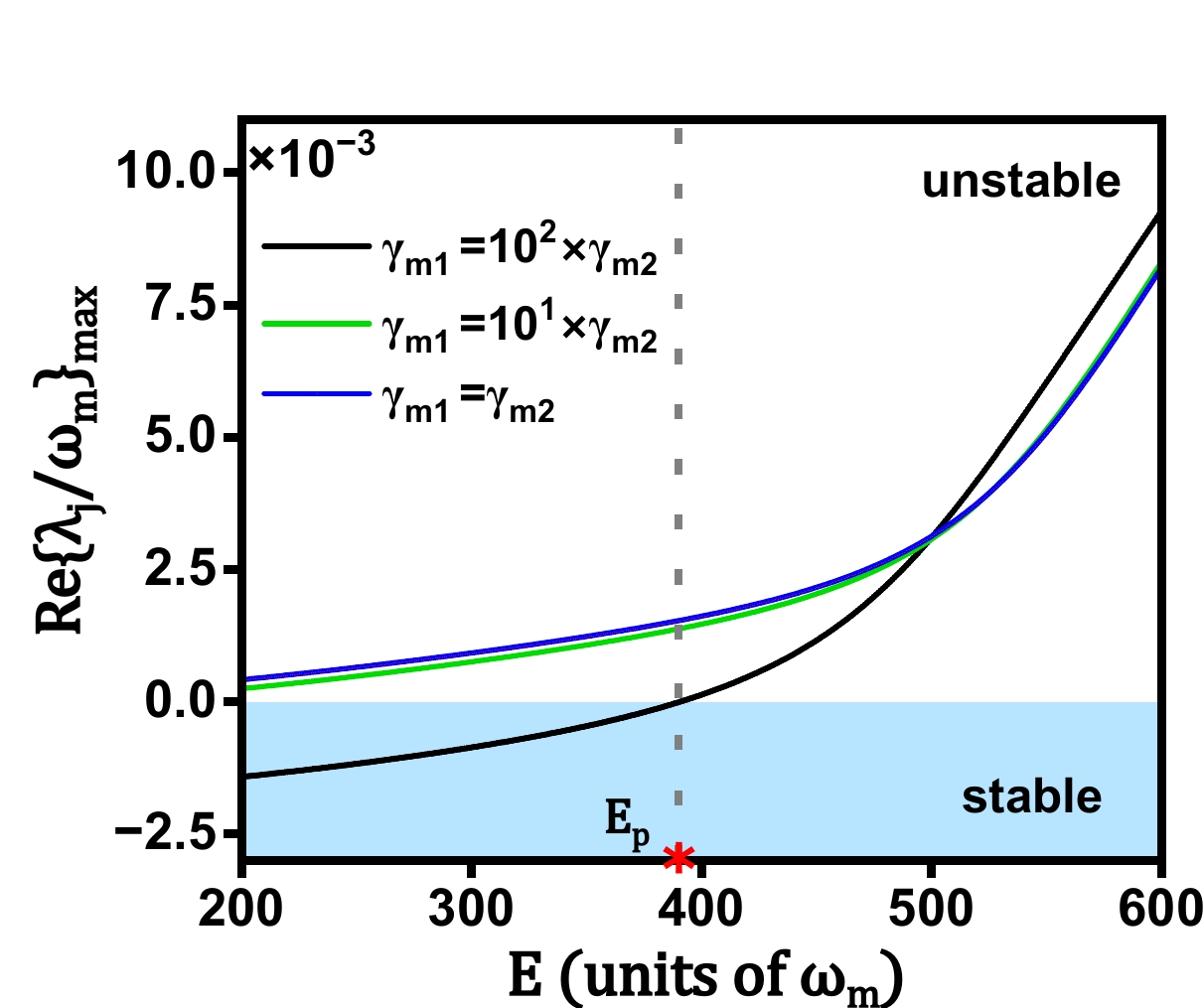}
\caption{(Color online) The maximum eigenvalues of $\mathcal{A}$ against the driving power strength for different damping ratios of the oscillators, other parameters remain the same. The critical driving power $E_p$ is depicted by the red star on the horizontal axis.}
{\label{fig3} }
\end{figure}

The stability conditions of the system are obtained by numerically solving the eigenvalues of the drift matrix $\mathcal{A}$, in which the system becomes unstable when any one of the real part of the eigenvalues becomes positive. Fig.\ref{fig3} shows the stable and unstable regions for varying driving strength, where we notice that for higher $\gamma_{m1}/\gamma_{m2}$ ratio, the system becomes unstable in the effective weak coupling regime ($E\gtrsim E_p$). On the other hand, fixing $\gamma_{m1}\sim\gamma_{m2}$ makes the system unstable throughout the whole driving power range. This would hamper the effect of EP on the emergence of entanglement and synchronization dynamics, as finite oscillations may get induced below the critical driving power $E<E_p$.

Since the fluctuation dynamics in Langevin equations are linearized and noises are also taken as zero-mean Gaussian distribution, the evolved states are also time-dependent Gaussian states with zero means irrespective of initial conditions \cite{chen2014enhancement}. Therefore, Gaussian dynamics can be fully characterized by the covariance matrix (CM) formalism \cite{RevModPhys.77.513}. Let, $\mathcal{V}$ be the covariance matrix whose elements are defined as
\begin{equation}
    \mathcal{V}=\frac{1}{2}\langle u_i(t) u_j(t) + u_j(t) u_i(t)\rangle
    \label{eq:8}
\end{equation}

Here $u_{j}$ is the $j^{th}$ entry of the quadrature vector $u$ defined and the evolution of the covariance matrix and its elements are governed by the following differential equation
\begin{equation}
    \partial_t\mathcal{V} = \mathcal{A}\mathcal{V} + \mathcal{V}\mathcal{A}^T + \mathcal{N}
    \label{eq:9}
\end{equation}

$\mathcal{N}$ is the diffusion matrix for noise, which satisfies the correlation formula
$\frac{1}{2}\langle n_i(t)n_j(t')+n_j(t')n_i(t)\rangle=N_{ij}\delta(t-t')$. This is used to deduce noise correlation vector as $\mathcal{N}=$Diag$[0,\gamma(2\bar{n}_{m}+1),\kappa,\kappa,0,\gamma(2\bar{n}_{m}+1),\kappa,\kappa]$. CM for the whole system has the following form
\begin{equation}
    \mathcal{V}_{8\times8}=
    \begin{pmatrix}
        \mathcal{V}_{m_1} & \mathcal{V}_{m_1,a_1} & \mathcal{V}_{m_1,m_2} & \mathcal{V}_{m_1,a_2}\\
        \mathcal{V}_{a_1,m_1} & \mathcal{V}_{a_1} & \mathcal{V}_{a_1,m_2} & \mathcal{V}_{a_1,a_2}\\
        \mathcal{V}_{m_2,m_1} & \mathcal{V}_{m_2,a_1} & \mathcal{V}_{m_2} & \mathcal{V}_{m_2,a_2}\\
        \mathcal{V}_{a_2,m_1} & \mathcal{V}_{a_2,a_1} & \mathcal{V}_{a_2,m_2} & \mathcal{V}_{a_2}
    \end{pmatrix}
    \label{eq:10}
\end{equation}

  Here $m_{1}$ and $m_{2}$ denote the mechanical modes of the vibrating oscillators and $a_{1}$ and $a_{2}$ are the modes corresponding to optical fields. Each block of $\mathcal{V}$ represents a $2\times 2$ square matrix. The off-diagonal elements of the matrix represent the covariance of different subsystems while diagonal elements refer variance of each system. From this matrix, we can easily calculate the correlation properties between two different subsystems. The coupled mechanical system can be easily described by extracting the submatrix $\mathcal{V'}$ from Eq.(\ref{eq:10}), which has the following form

  \begin{equation}
     \mathcal{V'}_{4\times4}=
    \begin{pmatrix}
        \mathcal{V}_{m_1} & \mathcal{V}_{m_1,m_2}\\
        \mathcal{V}_{m_1,m_2}^{T} & \mathcal{V}_{m_2}
     \end{pmatrix}
    \label{eq:11}
\end{equation}
By singular value decomposition, it can be shown that the $2\times 2$ symplectic matrices of Eq.(\ref{eq:11}) can be written as $\mathcal{V}_{mj}=(2\bar{n}_{m}+1)R(\phi)S(2r)R^{T}(\phi)$ where $S(r)= \exp[r(b_j^2-b_j^{\dagger 2})]$ is the squeezing operator for $j^{th}$ mechanical mode with squeezing parameter $r$ and $R(\phi)=\begin{pmatrix}
        \cos\phi_j & -\sin\phi_j\\
        \sin\phi_j & \cos\phi_j
    \end{pmatrix}$ is the phase rotation operator with rotation angle $\phi_j$.
    
\subsection{Quantum phase synchronization}
 In the previous section, we demonstrated that the oscillators are phase-locked for limit cycles. So the measurement of quantum phase synchronization is taken corresponding to the quadrature fluctuations of the operators, which can be evaluated by the position and momentum quadrature errors for the oscillators, given as $\delta q_- = \frac{1}{\sqrt{2}}[\delta q_1(t)-\delta q_2(t)]$ and $\delta p_- = \frac{1}{\sqrt{2}}[\delta p_1(t)-\delta p_2(t)]$. The figure of merit to calculate quantum phase synchronization is defined as \cite{PhysRevLett.111.103605}
\begin{equation}
    S_p(t) = \frac{1}{2}\langle\delta p'_{-}(t)^2\rangle^{-1}
    \label{eq:12}
\end{equation}
Here $\delta p'_{-}=\frac{1}{\sqrt{2}}[\delta p'_1(t)-\delta p'_2(t)]$ is the phase-locking operator associated with the mechanical oscillators.
Where
\begin{equation}
    \begin{pmatrix}
       \delta p'_{j}(t)\\
       \delta q'_{j}(t)
    \end{pmatrix}
    =R(\phi) \begin{pmatrix}
      \delta  p_j(t)\\
      \delta  q_j(t)
    \end{pmatrix}
    \label{eq:13}
\end{equation}
$R(\phi)$ is the rotation matrix and the phase is defined as $\phi_j=\arctan[\langle p_j\rangle/\langle q_j\rangle]\in[0,2\pi]$. In the case of quantum phase synchronization, we obtain equal quadrature variances for both oscillators i.e. $\langle \delta q'_{-}(t)^2 \rangle=\langle\delta p'_{-}(t)^2\rangle$ \cite{qiao2020quantum}. Analytical expression of $S_p$ can be easily obtained from the covariance matrix.

\subsection{Bipartite Gaussian entanglement}

The bipartite entanglement between the Gaussian states can be estimated by the following expression of logarithmic negativity \cite{PhysRevA.65.032314, PhysRevA.67.052315}
\begin{equation}
    E_n = \max[0,-\log(2\nu^{-})]
    \label{eq:14}
\end{equation}
where
\begin{equation}
    \nu^{-}= \sqrt{\frac{\Sigma-\sqrt{\Sigma^2 - 4\det(\mathcal{V'})}}{2}}
    \label{eq:15}
\end{equation}

is the smallest symplectic eigenvalue of the partial transpose of the submatrix $\mathcal{V'}$ in with
$\Sigma= \det(\mathcal{V}_{m_1})+\det(\mathcal{V}_{m_2})-2\det(\mathcal{V}_{m_1,m_2})$.
According to Simon's criterion of positive partial transpose (PPT) \cite{simon2000peres}, the necessary and sufficient condition for bipartite Gaussian states to be entangled is $\nu^{-}<0.5$.

\section{Results and discussion}

In this section, we establish the relationship between entanglement and phase synchronization between the coupled mechanical oscillators and discuss the significance of the EP in the proposed system for developing stable quantum correlation dynamics.

\subsection{Phase synchronization and entanglement dynamics}

\begin{figure}[htbp]
\includegraphics[width=0.85\linewidth]{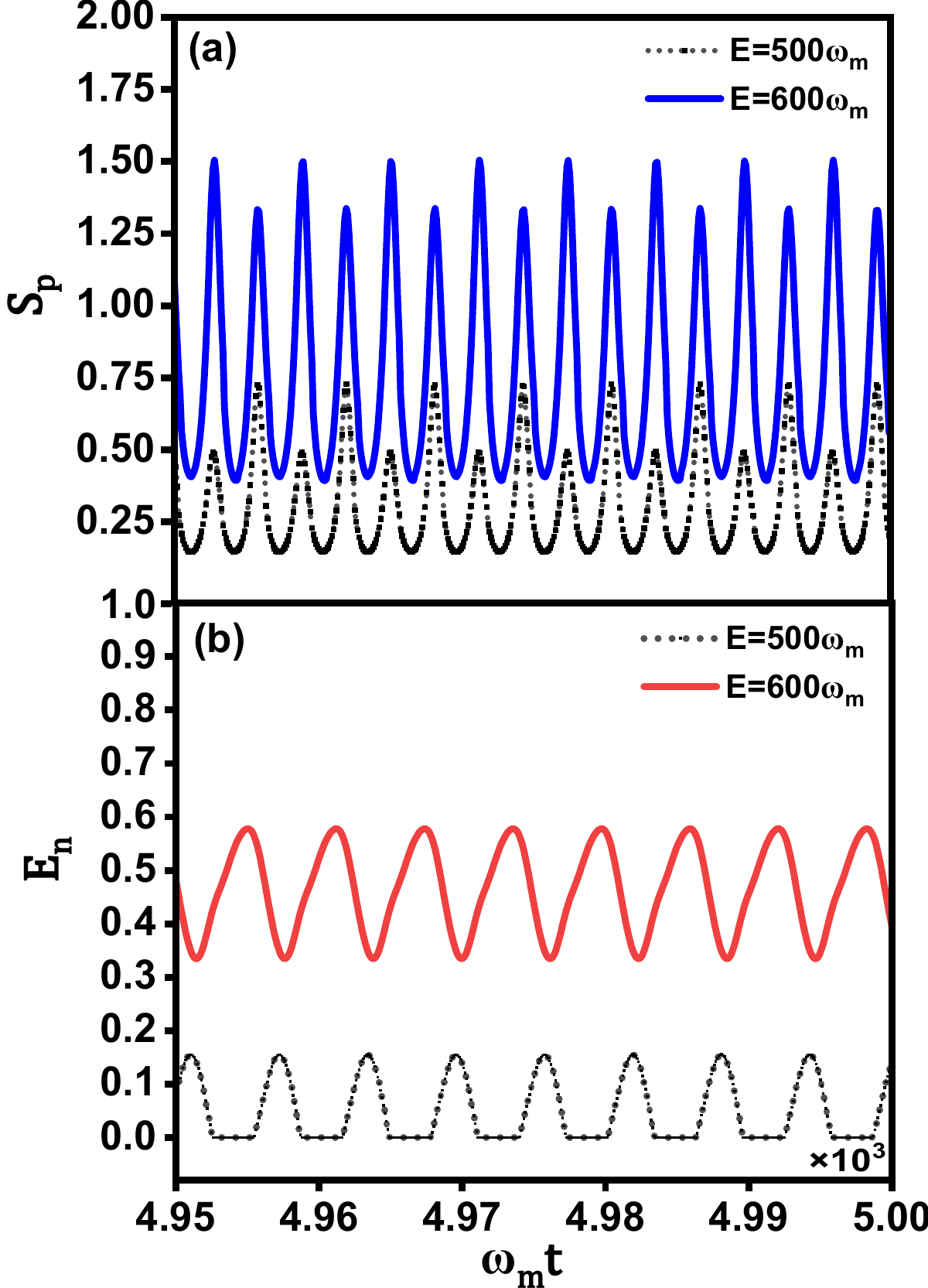}
\caption{ (Color online) (a) Quantum phase synchronization and (b) Entanglement dynamics of the two mechanical oscillators in the self-sustained regime with driving power strength $E=500\omega_m$ (dotted line) and $E=600\omega_m$ (solid line), other parameters remain same as used before.}
{\label{fig4} }
\end{figure}

\begin{figure*}[htbp]
\includegraphics[width=\linewidth]{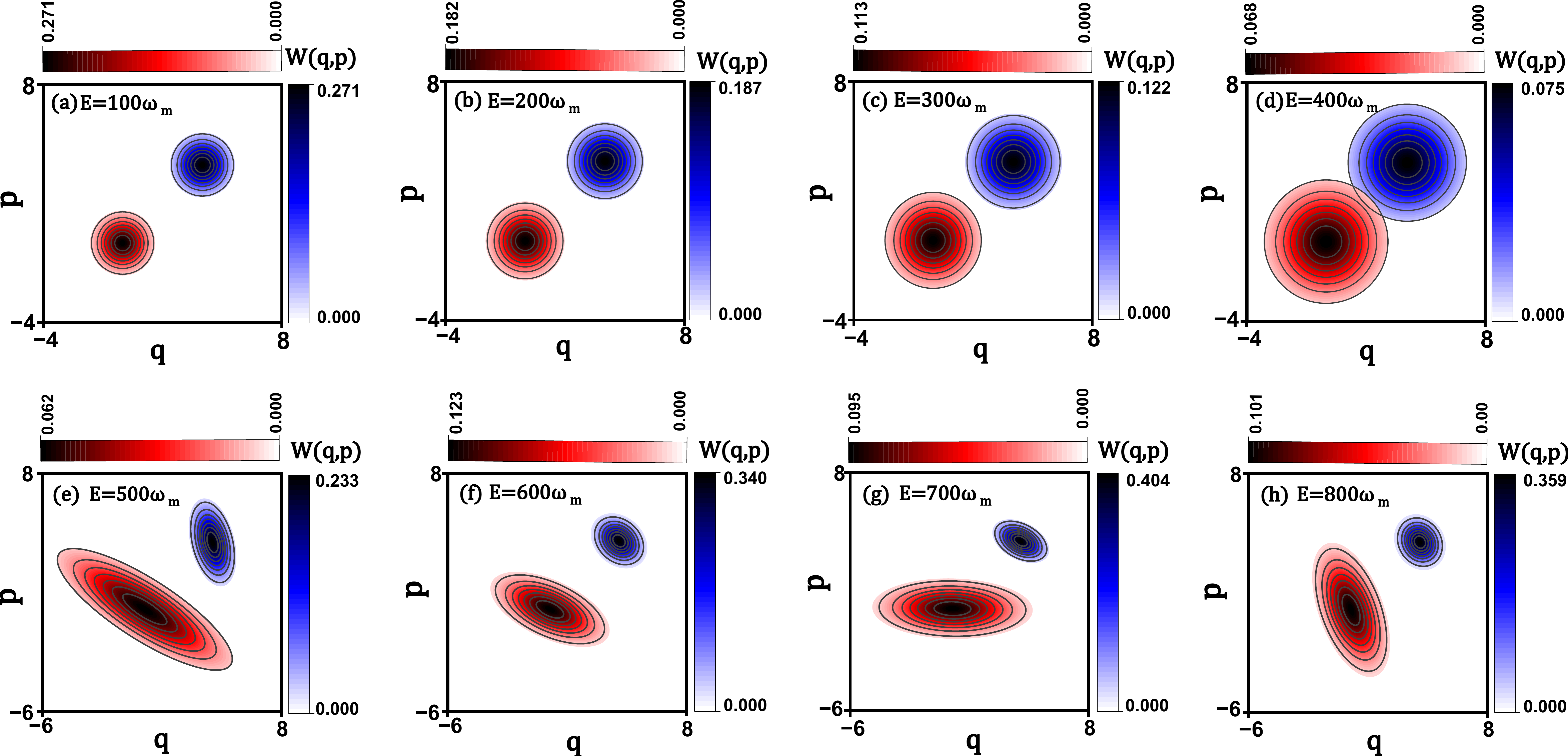}
\caption{(Color online) The Wigner distribution function $W(q,p)$ of gain (blue) and loss (red) oscillators at different driving powers. The upper panel (a)-(d) represent $W(q,p)$ upto exceptional point with $E=100\omega_m$, $E=200\omega_m$, $E=300\omega_m$, and $E=400\omega_m$, whereas in the lower panel (e)-(h) represent $W(q,p)$ after the exceptional point with $E=500\omega_m$, $E=600\omega_m$, $E=700\omega_m$, and $E=800\omega_m$, respectively. Other parameters remain the same, and time is fixed at $t/\tau=5000$ (where $\tau=1/\omega_m$) for both oscillators.}
\label{fig5}
\end{figure*}
We begin the analysis by numerically solving Eq.(\ref{eq:9}), which describes the behavior of the CM elements associated with the optical and mechanical modes. For the numerical simulation, the initial condition of $\mathcal{V}(0)=\frac{1}{2}$Diag$[1,1,1,1,1,1,1,1]$ is used. This corresponds to the vacuum state for both the cavities and thermal state for the mechanical oscillators with mean thermal phonon number $\bar{n}_m=0$, which can be achieved by pre-cooling them to their ground state \cite{chakraborty2018entanglement}.
The CM specifies how quadrature fluctuations are correlated across different bipartite subsystems, and its elements are specified in Eq.(\ref{eq:10}). From the matrix $\mathcal{V}$, we have extracted $\mathcal{V'}$ as given in Eq.(\ref{eq:11}), which only contains information about the vibrating oscillators. $\mathcal{V'}$ is associated with the characterization of phase synchronization ($S_p$) and entanglement ($E_n$) parameters, as defined in Eq.(\ref{eq:12}) and (\ref{eq:14}). 

While tuning the driving power $E$ and surpassing the exceptional point threshold ($E>E_p$), the coupled oscillators enter into the limit cycle regime, where the consistent correlation of the quadrature fluctuations becomes noticeable. Beyond the EP zone, indicated by the shaded area in Fig.\ref{fig2}(d), there is a sudden surge in mechanical vibrations, triggering instability within the system. In this case, the EP corresponds to critical driving power $E_p=390\omega_m$, however, quantum correlation dynamics did not prevail until $E$ reaches $490\omega_m$, which is in congruence with the earlier finding \cite{tchodimou2017distant}. The steady dynamics of the phase synchronization parameter $S_p$ can be observed from Fig.\ref{fig4}(a) for two different power levels, $E=500\omega_m$ and $E=600\omega_m$ in the limit cycle regime. $S_p$ increases with high driving powers, as the optomechanical nonlinearity becomes strong in the cavities. A similar behavior is also observed for entanglement dynamics $E_n$, shown in Fig.\ref{fig4}(b). $E_n$ depicts the dynamics of logarithmic negativity ($E_n>0$) for the same driving powers as considered for the synchronization parameter $S_p$. The enhancement in $S_p$ and $E_n$ occurs as long as the driving power compensates for the weak effective coupling condition. Both dynamics can be further enhanced by increasing the driving power, which effectively increases the nonlinearity of the system. We also notice the death and rebirth of entanglement \cite{PhysRevA.81.052330} happens at the lower driving strength ($E=500\omega_m$), which is near the EP and vanishes quickly as the power increases. This type of dynamics changes differently with higher power levels beyond $E\gg E_p$ and different frequency mismatches between the oscillators, which we will discuss later in this section.
Over time, entanglement and synchronization both exhibit periodic variations. This is due to the fact that quantum fluctuations follow classical orbits as long as all Lyapunov exponents of the classical equations are negative \cite{meng2020quantum}.
It is essential to understand that steady quantum correlation dynamics arise with weak coupling conditions only. When the driving amplitude is not strong enough ($E<E_p$), classical dynamics cannot be sustained, and the oscillators do not entangle or synchronize. 

\subsection{Wigner distribution and fidelity}
In order to further confirm the influence of exceptional points on synchronization and entanglement, we plot the two-mode Wigner distribution function $W(q,p)$ of the coupled oscillators for various driving powers. The Gaussian Wigner distributions of the two mechanical modes $m_1$ and $m_2$ are defined as \cite{olivares2012quantum, weedbrook2012gaussian}
\begin{equation}
    W(q,p)=\frac{1}{2\pi\sqrt{\det(\mathcal{V}_{m_j})}}\exp[-\frac{u_j \mathcal{V}^{-1}_{m_j} u^{T}_j }{2}]
    \label{eq:16}
\end{equation}
\begin{figure*}[htbp]
\includegraphics[width=\linewidth]{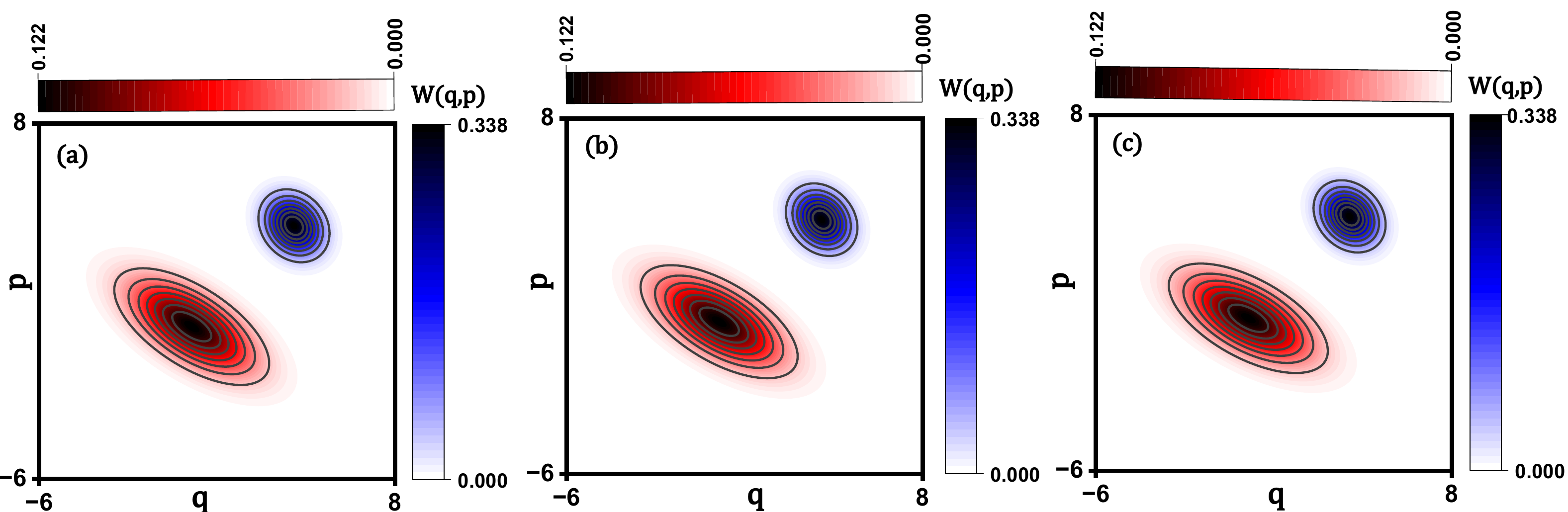}
\caption{(Color online) Time evolution of the Wigner distributions of the gain (loss) oscillator with driving power $E=600\omega_m$ at different times (a) $t/\tau=3000$, (b) $t/\tau=4000$ and (c) $t/\tau=5000$, ($\tau=1/\omega_m$) other parameters remain the same.}
\label{fig6}
\end{figure*}
where $j=1,2$ represents the two coupled oscillators and $u_j$ and $\mathcal{V}_{mj}$ are first and second-order moment vectors of the $j^{th}$ mechanical mode.  The first-order moment vector, $u_j$, indicates the position of the origin and the second-moment vector $\mathcal{V}_{mj}$ can be found from diagonal block matrices of Eq.(\ref{eq:11}). However, $u_j$ does not provide any relevant information and can be conveniently set to zero. The phase space dynamics of the gain (loss) oscillators are represented with blue (red) colors in Fig.\ref{fig5}. The driving power $E=100\omega_m$ corresponds to both oscillators being in their ground state with nearly equal net gain or loss, as shown in Fig.\ref{fig5}(a). This occurs because of the strong coupling, which causes a coherent exchange of energy between them. As a result, there is no indication of squeezing or rotation in the dynamics of the phase space. With the rise in driving power, (b) $E=200\omega_m$ and (c) $E=300\omega_m$, the dispersion in phase space causes a decrease in the Wigner density function. This fact can also be verified by the decaying dynamics of classical oscillation in Fig.\ref{fig2}(a). Near the exceptional point at $E=400\omega_m$ in Fig.\ref{fig5}(d), we notice an overlapping in the Wigner distribution functions and the two mechanical modes are closest. Due
to mechanical amplification at EP, the Wigner functions become delocalized, and abrupt stretching occurs in phase space, which is a sign of dynamical instability \cite{chakraborty2018entanglement}. But this delocalization vanishes quickly as the system moves away from the EP zone. From Fig.\ref{fig5}(e), the squeezing effect is evident in both oscillators, indicating entanglement. Additionally, Wigner's distributions start to rotate in phase space because the weak coupling cannot support energy exchange anymore and mechanical energy is localized in each oscillator. As we move further from EP, the degree of phase space rotation changes, as well as the squeezing effect. The origin of phase synchronization dynamics and entanglement of the limit cycle power levels in Fig.\ref{fig4}(a),(b) can be traced back to this point.
Fig.\ref{fig5}(f)-(h), represents several other Wigner distributions at power levels $E=600\omega_m$, $E=700\omega_m$ and $E=800\omega_m$. Also, the shape and phase space rotation angle of Wigner functions remain constant over time, at a fixed driving power. Fig.\ref{fig6} represents $W(q,p)$ at three subsequent times with power level remaining $E=600\omega_m$. It can be noted that when two oscillators are phase synchronized, their angle of rotation remains constant and the squeezing parameter does not vary with time. Therefore, the proposed system exhibits both phase synchronization and entanglement simultaneously.

\begin{figure}[htbp]
\includegraphics[width=\linewidth]{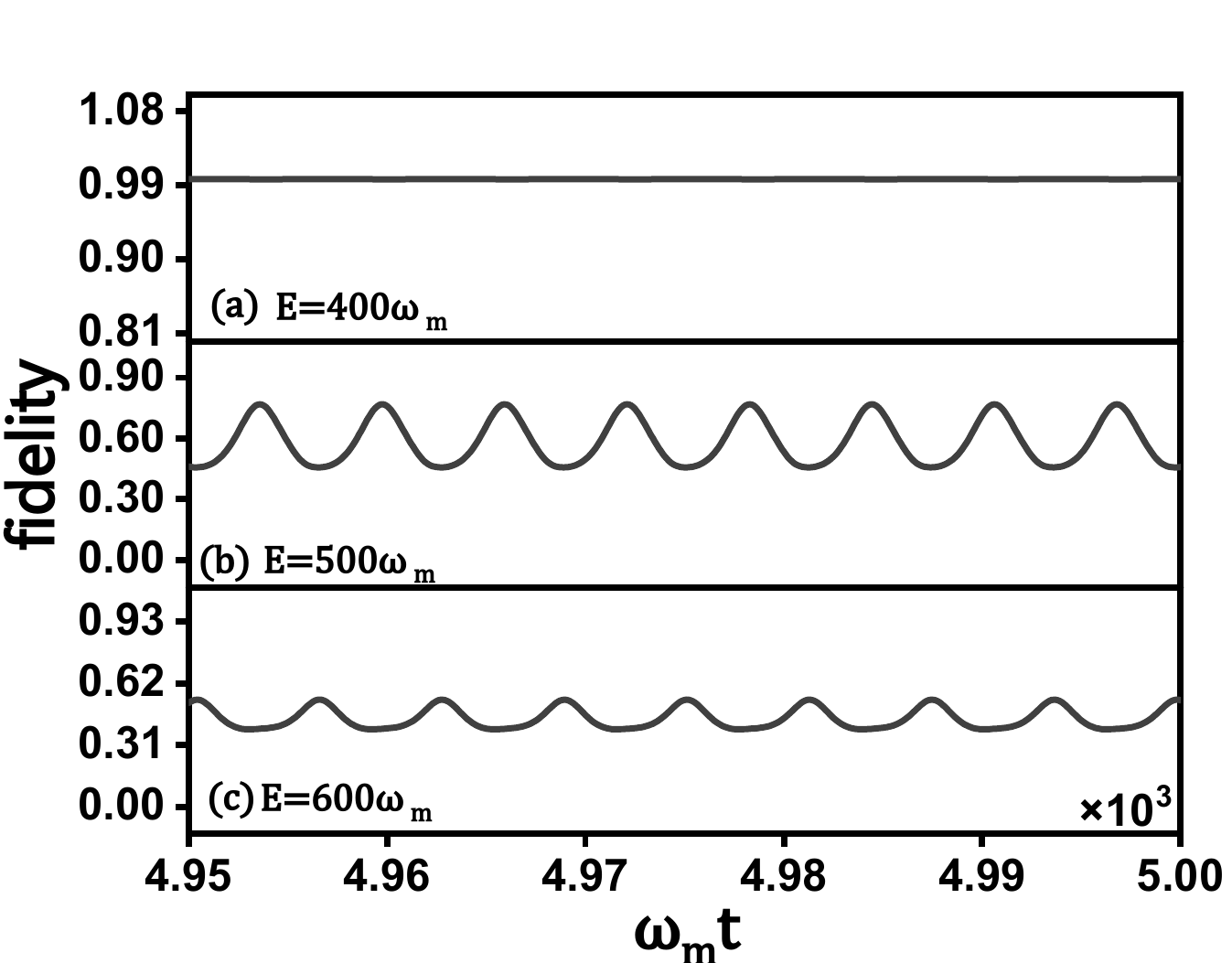}
\caption{Fidelity, $f$, of the coupled mechanical systems near the exceptional point (a) $E=400\omega_m$ and after the EP junction at (b) $E=500\omega_m$ and (c) $E=600\omega_m$, other parameters remain the same. }
\label{fig7}
\end{figure}

Another important aspect is to verify the behavior of Wigner functions by calculating fidelity for the Gaussian states. In this system, fidelity is determined by comparing the overlap of two Gaussian states. Essentially, it measures the level of similarity between these states, defined as \cite{weedbrook2012gaussian, olivares2012quantum}
\begin{equation}
    f=\frac{\exp[-\frac{1}{2}(u_1-u_2)(\mathcal{V}_{m_1}+\mathcal{V}_{m_1})^{-1}(u_1-u_2)^{T}]}{\sqrt{\Delta+\delta}-\sqrt\delta}
    \label{eq:17}
\end{equation}
With $\Delta=\det(\mathcal{V}_{m_1}+\mathcal{V}_{m_1})$ and $\delta=4(\det[\mathcal{V}_{m_1}]-0.25)(\det[\mathcal{V}_{m_2}]-0.25)$. The dynamics of fidelity $f$ is shown in Fig.\ref{fig7}. As expected from the Wigner distribution functions of Fig.\ref{fig5}(d), fidelity near the exceptional point i.e. $E=400\omega_m$ is almost unity. When we move further from EP, fidelity decreases, which is also evident from the shape of Wigner distributions. Fig.\ref{fig7} represents fidelity dynamics at two limit cycle power levels (b) $E=500\omega_m$ and (c) $E=600\omega_m$, same as used before. As fidelity fluctuates at high power, there is a greater likelihood of phase synchronization and entanglement. The fidelity dynamics indicate two initially uncorrelated Gaussian states were squeezed and phase space rotated by surpassing the EP, if they were unchanged, we would not observe these phenomena.

\subsection{Effect of frequency mismatch and finite thermal phonons}

\begin{figure}[htbp]
\includegraphics[width=\linewidth]{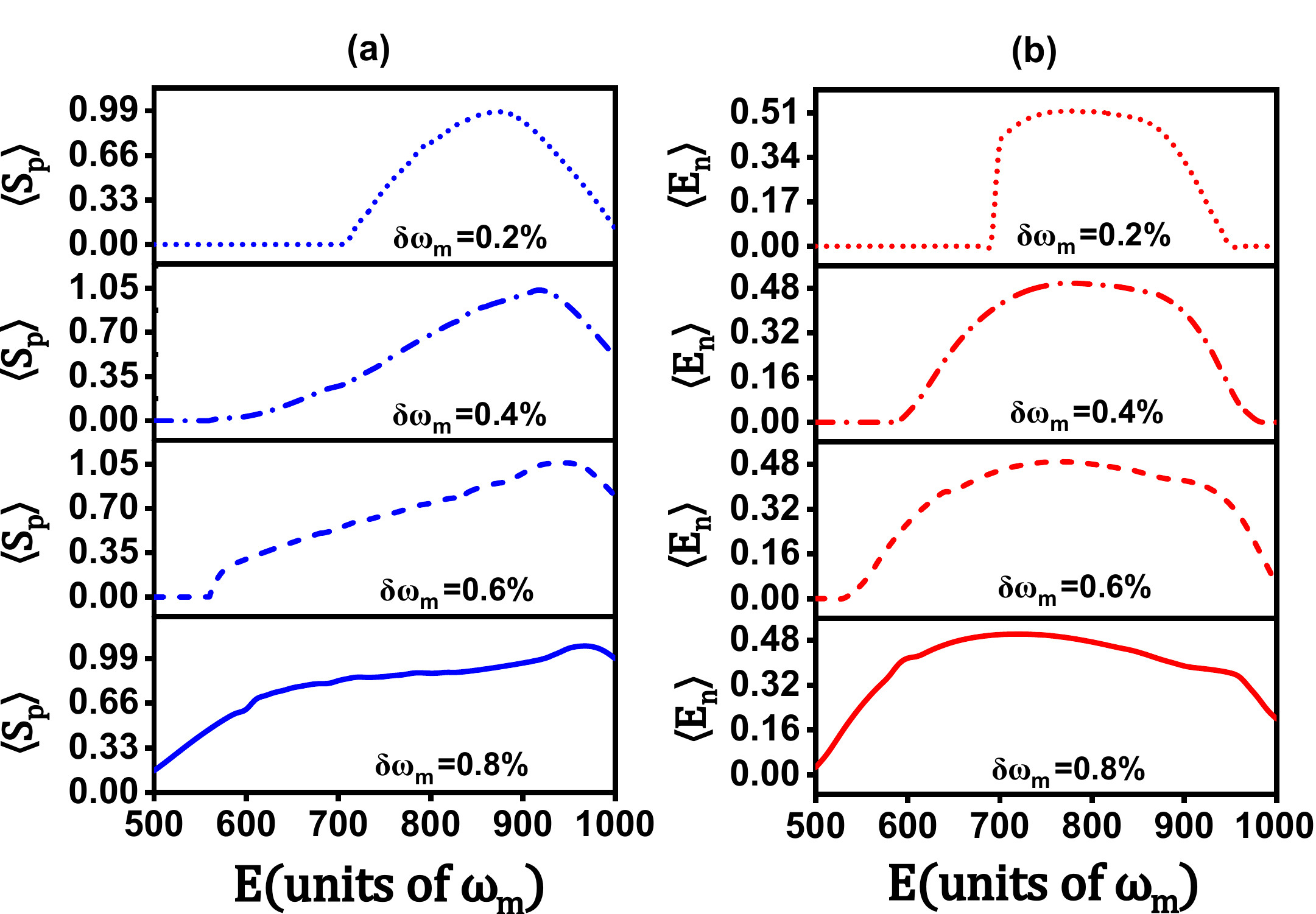}
\caption{ (Color online) Time average of (a) phase synchronization ($S_p$) and (b) entanglement ($E_n$) parameters for different frequency mismatch $\delta\omega_m$ with driving power after EP region, the other parameters are same as used.}
\label{fig8}
\end{figure}

It is important to consider the frequency difference between the coupled oscillators when examining phase synchronization and entanglement characteristics. To explore the deviations, the frequency mismatch is set with four distinct values in the range $0-1\%$ of $\omega_m$. Note that $\delta\omega_m(=\omega_{m2}-\omega_{m1})$ should be maintained very small due to the assumption considered in the eigenvalue Eq.(\ref{fig6}) in the classical analysis. Fig.\ref{fig8} shows the driving strengths $E$ in the limit cycle regime after the EP-induced amplification, where the time average of the quantum phase synchronization parameter (a) $S_p$ and the entanglement parameter logarithmic negativity (b) $E_n$ is plotted while varying the driving powers. The time averages of the parameters are calculated by using the formula $\langle h(t) \rangle =\lim_{t\to\infty}\frac{1}{T}\int_{0}^{T}h(t)dt$, where $h(t)=S_p, E_n$. It is clear from Fig.\ref{fig8} that the maximum of the averages of $S_p$ and $E_n$ occurs when the frequency mismatch is smallest. i.e. $\delta\omega_m=0.2\%$. Interestingly, we observe a greater tendency towards synchronization and entanglement with an increase in $\delta\omega_m$, contradictory to the classical case,  resulting in similarity to the blockade phenomenon \cite{PhysRevA.97.013811, liao2019quantum}. When the deviation in frequency, $\delta\omega_m$, is 0.8\%, the maximum range of driving power in the limit cycle region is required to synchronize and entangle the oscillators. 
The quantum correlation dynamics in both cases show a decline after reaching a maximum value as higher driving strength $E\gg E_p$ creates a significant difference in oscillation amplitudes shown in Fig.\ref{fig2}(d). Although phase synchronization lasts longer at higher power than entanglement, it shows that entanglement is more sensitive to frequency differences and stronger driving forces.

\begin{figure}[htbp]
\includegraphics[width=\linewidth]{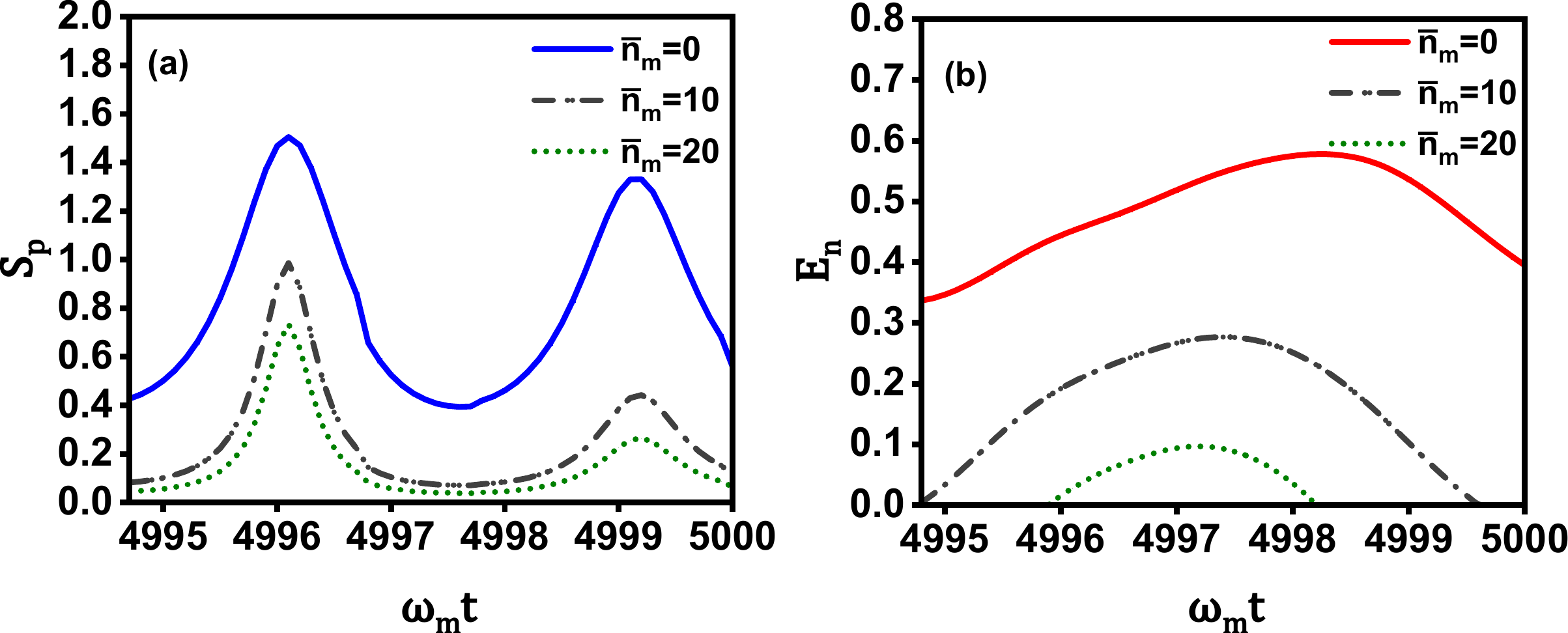}
\caption{(Color online) Time evolution of (a) phase synchronization and (b) entanglement at different mean thermal phonon numbers such as $\bar{n}_m=0$, $\bar{n}_m=10$ and $\bar{n}_m=20$ at driving power $E=600\omega_m$, other parameters same as above.}
\label{fig9}
\end{figure}

The impact of thermal phonons on quantum dynamics is another essential parameter to analyze the deviations. The calculations mentioned so far do not account for thermal noise. However, as the system temperature rises, there is a corresponding increase in mean thermal phonon numbers. Fig.\ref{fig9} displayed dynamics of phase synchronization and entanglement for various thermal phonon numbers when increased from idealistic condition of $\bar{n}_m=0$ to $\bar{n}_m=10$ and to $\bar{n}_m=20$. As the temperature increases, the amplitudes of entanglement and phase synchronization dynamics diminish, due to the decoherence. However, phase synchronization is more resilient than entanglement, even at high temperatures.

\section{Conclusion}

In this paper, we explored entanglement and quantum phase synchronization dynamics in a gain-loss optomechanical system with mechanically coupled oscillators. By applying opposite detunings, we induced gain or loss effects in the mechanical oscillators. Using experimentally feasible parameters, we observed various oscillation dynamics classically, including damping and self-sustained vibrations. These oscillators showed phase-locked behavior in the weak coupling limit cycle regimes, which could be accessed by adjusting laser power and tuning the exceptional point. Quantum correlations of quadrature fluctuation operators emerged during limit cycle oscillations, revealing entanglement and synchronized quantum phases between the coupled mechanical modes. As driving power increased, effective coupling became weaker, and entanglement and synchronization dynamics enhanced. These phenomena initially grew but later decreased due to factors like higher driving strength and various frequency mismatches. Corresponding Wigner function distributions help to visualize the evolved Gaussian states that are squeezed and phase-space rotated. Remarkably, phase synchronization remained robust against thermal noise, while entanglement was more sensitive to temperature changes. Our numerical calculations have shown that mechanical oscillators can manipulate Gaussian quantum information by adjusting the EP through the phonon transfer mechanism acting as Gaussian channels.

\appendix
\section{Drift Matrix}
The drift matrix $\mathcal{A}$ is given by
\begin{widetext}
\begin{equation}
\mathcal{A}(t)=
  \setlength{\arraycolsep}{2pt}
  \renewcommand{\arraystretch}{0.6}
  \begin{pmatrix}
    
0 & \omega_{m1} & 0 & 0 & 0 & 0 & 0 & 0\\
 -\omega_{m1} & -\gamma_1 & \sqrt{2}g_0 Re(\langle a_1\rangle) & \sqrt{2}g_0 Im(\langle a_1\rangle) & J  & 0 & 0 & 0\\
 \sqrt{2}g_0 Im(\langle a_1\rangle) & 0 & -\kappa & -(\Delta_1+g_0\langle q_1\rangle) & 0 & 0 & 0 & 0\\
   \sqrt{2}g_0 Re(\langle a_1\rangle) & 0 & (\Delta_1+g_0\langle q_1\rangle) & -\kappa & 0 & 0 & 0 & 0\\
  0 & 0 & 0 & 0 & 0 & \omega_{m2} & 0 & 0\\
 J & 0 & 0 & 0 & -\omega_{m2} & -\gamma_2 & \sqrt{2}g_0 Re(\langle a_2\rangle) & \sqrt{2}g_0 Im(\langle a_2\rangle)\\
 0 & 0 & 0 & 0 &  \sqrt{2}g_0 Im(\langle a_2\rangle) & 0 & -\kappa & -(\Delta_2+g_0\langle q_2\rangle)\\
 0 & 0 & 0 & 0 &  \sqrt{2}g_0 Re(\langle a_2\rangle) & 0 & (\Delta_2+g_0\langle q_2\rangle) & -\kappa\\
  \end{pmatrix}
  \nonumber
\end{equation}
\end{widetext}

\nocite{*}

\bibliography{apssamp}

\end{document}